# 3D Simulation of Nanowire FETs using Quantum Models

Vijay Sai Patnaik, Ankit Gheedia and M. Jagadesh Kumar

Abstract— After more than 30 years of validation of Moore's law, the CMOS technology has already entered the nanoscale (sub-100nm) regime and faces strong limitations. The nanowire transistor is one candidate which has the potential to overcome the problems caused by short channel effects in SOI MOSFETs and has gained significant attention from both device and circuit developers. In addition to the effective suppression of short channel effects due to the improved gate strength, the multi-gate NWFETs show excellent current drive and have the merit that they are compatible with conventional CMOS processes. To simulate these devices, accurate modeling and calculations based on quantum mechanics are necessary to assess their performance limits, since cross-sections of the multigate NWFETs are expected to be a few nanometers wide in their ultimate scaling. In this paper we have explored the use of ATLAS including the Bohm Quantum Potential (BQP) for simulating and studying the short-channel behaviour of nanowire FETs.

## Introduction

Some of the fundamental problems of MOSFET-inspired devices for sub-10nm channel length are expected to be electrostatic limits, source-to-drain tunneling, carrier mobility, process variations, and static leakage. A simultaneous concern is that of power scaling. Nevertheless. it appears that emerging device architectures can extend the CMOS lifetime and provide solutions to continue scaling into the nanometer range, or at least until the 10 nm wall is reached [1-3]. But what comes after this limit? The nanowire transistor is one candidate which has gained significant attention from both device and circuit developers because of its potential for building highly dense and high performance electronic products. Nanowire transistors can be made using different materials on low cost substrates such as glass or plastics. Si and Ge nanowire transistors are of particularly more importance because of their compatibility with CMOS technology[4-7].

The objective of this paper is to explain how ATLAS3D can be used to simulate silicon nanowire field effect transistors. We demonstrate that by using the Bohm Quantum Potential to include the confinement effects and by calibrating the BQP model against the 2D Schrodinger-Poisson simulation, the scaling behaviour of silicon nanowire FETs can be successfully predicted.

*The authors are with the Department of Electrical Engineering, Indian Institute of Technology, Huaz Khas, New Delhi 110 016, India (Email: vijaysai.patnaik@gmail.com)*

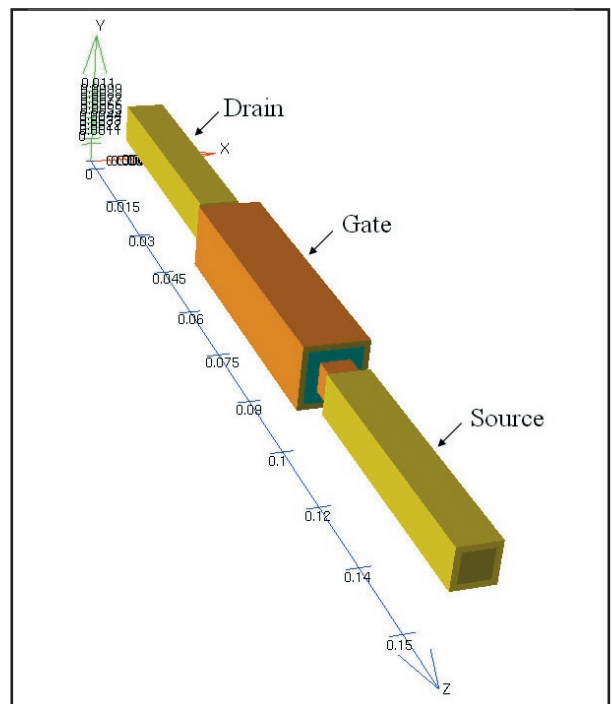

Fig.1 3D Nanowire FET generated using ATLAS.

## Quantum Models in ATLAS3D

Quantum3D provides a set of models for simulation of various effects of quantum confinement and quantum transport of carriers in semiconductor devices such as the silicon nanowire FETs shown in Fig. 1. A self-consistent Schrödinger – Poisson solver allows calculation of bound state energies and associated carrier wave function self consistently with electrostatic potential.

### A. Self-consistent Coupled Schrodinger-Poisson Model [8]

To model the effects of quantum confinement, ATLAS also allows the solution of Schrödinger's Equation along with the fundamental device equations. The solution of Schrödinger's Equation gives a quantized description of the density of states in the presence of quantum mechanical confining potential variations.

The SCHRO parameter of the MODEL statement enables the self-consistent coupled Schrödinger-Poisson model for electrons. The P.SCHRO parameter in the MODEL statement enables the Schrödinger-Poisson model for holes. ATLAS2D can solve Schrödinger equation in 1D slices or 2D plane. In the case of cylindrical coordinates, Schrödinger equation is solved in radial direction for different orbital quantum numbers and for all slices perpendicular to the axis. By default, a 1D equation is solved. To enable the 2D model, we need to specify the option 2DXY.SCHRO on the MODEL statement. ATLAS3D solves a 2D Schrödinger equation in y-z slices, perpendicular to x axis and the option 2DXY.SCHRO is optional.

When the quantum confinement is in one dimension (along y-axis), the calculation of the quantum electron density relies upon a solution of a 1D Schrödinger equation solved for eigen state energies $E_{iv}(x)$ and wave-functions $\Psi_{iv}(x, y)$ at each slice perpendicular to x-axis and for each electron valley (or hole band) $v$ as described below:

$$-\frac{h^2}{2}\frac{\partial}{\partial y}\left(\frac{1}{m^v_y(x,y)}\frac{\partial \psi_{iv}}{\partial y}\right) + E_c(x,y)\psi_{iv} = E_{iv}\Psi_{iv} \quad (1)$$

Here, (x, y) is a spatially dependent effective mass in y-direction for the $v$-th valley and $E_C(x, y)$ is a conduction band edge. The equation for holes is obtained by substituting hole effective masses instead of electron ones and valence band edge $-E_V(x,y)$ instead of $E_C(x, y)$. Analogously, in cylindrical coordinates, the equation for the radial part of the wave function $R_{im}v(r)$ for each orbital quantum number m reads as

$$-\frac{h^2}{2}\left[\frac{1}{r}\frac{\partial}{\partial y}\left(\frac{1}{m^v_y(r,z)}r\frac{\partial R_{imv}}{\partial r}\right) - \frac{1}{m^v_y(r,z)}\frac{m^2}{r^2}\right] + E_c(r,z)R_{imv} = E_{imv}R_{imv} \quad (2)$$

In the case where the mass is isotropic (i.e., mass is the same in all directions), only one solution to Schrödinger's equation is obtained with the appropriate mass. ATLAS will automatically determine the appropriate number of valleys based on the material in question. We can, however, limit the number of directions in k space by using the NUM.DIRECT. To choose the number of electron valleys, we need to specify the NUM.DIRECT parameter on the MODELS statement. If you set NUM.DIRECT to 1, we will obtain the solution for isotropic effective mass (MC on the MATERIAL statement). If we set NUM.DIRECT to 3, we obtain a solution for a single lateral effective mass and two transverse masses (ML, MT1 and MT2 on the MATERIAL statement). In a special case of a 1D confinement and equivalent transverse masses, we can set NUM.DIRECT to 2 and only two solutions will be obtained (ML and MT1) with appropriate degeneracy factors. To specify how many valence bands to consider, we need to specify the NUM.BAND parameter. If we set NUM.BAND to 1, you will obtain a Schrödinger solution for only one valence band (MV on the MATERIAL statement). If you set NUM.BAND to 3, you will cause solutions for heavy holes, light holes and holes in the split off band (MHH, MLH, and MSO on the MATERIAL statement).

Using Fermi-Dirac statistics, the discrete nature of the quantized density of states reduces the integral over energy to a sum over bound state energies. The expression for the electron concentration then becomes:

$$n(x,y) = 2^k \frac{B^r}{\pi h^2} \sum_v \sqrt{m_x^v(x,y)m_z^v(x,y)} \sum_{i=0}^{\infty} |\psi_{iv} \cdot (x,y)|^2 \ln[1+\exp(-\frac{E_{iv}-E_F}{k_B T})] \quad (3)$$

for 1D confinement and

$$n(x,y) = 2\frac{\sqrt{2k_B T}}{h} \sum_v \sqrt{m_z^v(x,y)} \cdot \sum_{i=0}^{\infty} |\psi_{iv}(x,y)|^2 F_{-1/2}(-\frac{E_{iv}-E_F}{k_B T}) \quad (4)$$

for a 2D confinement case and cylindrical case, where F-1/2 is the Fermi-Dirac integral of order -1/2.

ATLAS solves the one-dimensional Schrödinger's Equation along a series of slices in the y direction relative to the device. The location of the slices in the y direction is developed in two ways. For rectangular ATLAS-defined meshes, the slices will automatically be taken along the existing mesh lines in the ATLAS mesh. If the mesh is non-rectangular or not an ATLAS defined mesh or both, we must specify a rectangular mesh. To do this, we need to specify the locations of individual mesh lines and their local spacing using the SX.MESH and SY.MESH statements like that to the specification of a device mesh using the X.MESH and Y.MESH or a laser mesh using the LX.MESH and LY.MESH statements. The 2D Schrödinger equation in x-y plane can be solved on a general non-rectangular mesh. Specifying a rectangular mesh in addition to imported mesh is unnecessary.

Once the solution of Schrödinger's Equation is taken, carrier concentrations calculated from Equation (3) and Equation (4) are substituted into the charge part of Poisson's Equation. The potential derived from the solution of Poisson's Equation is substituted back into Schrödinger's Equation. This solution process (alternating between Schrödinger's and Poisson's equations) continues until convergence and a self-consistent solution of Schrödinger's and Poisson's equations is reached.

Since the wave functions diminish rapidly from the confining potential barriers in the Schrödinger solutions, the carrier concentrations become small and noisy. We can refine these carrier concentrations by setting a minimum carrier concentration using the QMINCONC parameter on the MODELS statement. This parameter sets the minimum carrier concentration passed along to the Poisson solver and the output to the structure files. The transition between the Schrödinger solution and the minimum concentration is refined between 10×QMINCONC and QMINCONC so that it is continuous in the first derivative.

We can use the SAVE statement or the OUTFILE parameter on the SOLVE statement to write the solutions of the self-consistent system into a structure file. These structure files will then contain the self-consistent potential and electron or hole concentrations. The Eigen energies and functions can also be written to the structure file by specifying the EIGENS parameter on the OUTPUT statement. This parameter specifies the number of Eigen energies/wave functions to be written.

The number of Eigen values solved is limited to a number of 2 less than the total number of grid points in the Y direction. Note that the self-consistent solution of Schrödinger's Equation with the Poisson's Equation doesn't allow solutions for the electron and hole continuity equations in the current ATLAS version. Non-self-consistent solutions, however, can be obtained by setting the POST.SCHRO parameter in the MODELS statement. These non-self-consistent solutions are obtained by solving Schrödinger's Equation only after convergence is obtained. That way, you can obtain Schrödinger solutions with the electron and hole continuity equations.

*B. Bohm Quantum Potential (BQP) [9]*

This model was developed for SILVACO by the University of Pisa and has been implemented into ATLAS with the collaboration of the University of Pisa. This is an alternative to the Density Gradient method and can be applied to a similar range of problems. There are two advantages to using Bohm Quantum Potential (BQP) over the density gradient method. First, it has better convergence properties in many situations. Second, you can calibrate it against results from the Schrödinger-Poisson equation under conditions of negligible current flow.

The model introduces a position dependent Quantum Potential, Q, which is added to the Potential Energy of a given carrier type. This quantum potential is derived using the Bohm interpretation of quantum mechanics and takes the following form

$$Q = -\frac{h^2}{2} \frac{\gamma \underline{\Delta}(M^{-1}\underline{\Delta}(n^\alpha))}{n^\alpha} \quad (5)$$

where $\gamma$ and $\alpha$ are two adjustable parameters, $M_{-1}$ is the inverse effective mass tensor and n is the electron (or hole) density. This result is similar to the expression for the quantum potential in the density gradient model with $\alpha = 0.5$, but there are some differences about how they are implemented.

The Bohm Quantum Potential (BQP) method can also be used for the Energy balance and hydrodynamic models, where the semi-classical potential is modified by the quantum potential the same way as for the continuity equations.

The iterative scheme used to solve the non-linear BQP equation along with a set of semi-classical equations is as follows. After an initial semi-classical solution has been obtained, the BQP equation is solved on its own Gummel iteration to give Q at every node in the device. The semi-classical potential is modified by the value of Q at every node and the set of semi-classical equations is then solved to convergence as usual (using a Newton or Block iterative scheme). Then, the BQP equation is solved to convergence again and the process is repeated until self-consistency is achieved between the solution of the BQP equation and the set of semi-classical equations. The set of semi-classical equations solved can be any of the combinations usually permitted by ATLAS.

To use the BQP model for electrons (or holes), we need to specify BQP.N (BQP.P) in the MODELS statement. We can also set the parameter values ($\alpha$ and $\gamma$) and the direction of the quantization (confinement).

*C. Calibration against Schrödinger-Poisson Model*

We can obtain close agreement between BQP and the results of Schrödinger-Poisson (S-P) calculations for any given class of device. ATLAS has a Schrödinger-Poisson model that can model spatial confinement in only one direction. Therefore, calibration is currently restricted to this case. To obtain comparisons with S-P results, ATLAS recommends to use either the new quasi-static capacitance-voltage profile feature or compare charge-voltage curves. This will ensure similar charge control properties between the two models.

The first part of the calibration is to choose a suitable biasing for the device. There should be negligible current flow and quantum confinement effects that manifest at the chosen biases. The second part of the calibration is to set the appropriate BQP parameters in the MATERIAL or MODELS statements, and to set CARRIERS=0 in the METHOD statement.

This will cause the BQP equation to be coupled with Poisson's equation using the charge density terms.

$$n = N_C \exp(-\frac{E_C + qQ}{kT}) \quad \text{and} \quad p = N_V \exp(-\frac{qQ - E_V}{kT}) \quad (6)$$

This gives the same results as solving the current continuity equations with the constraint of zero current density.

The third part of calibration is to choose the quantity to compare with S-P results. For example, for a MOSFET holding the drain and source voltages at the same bias and ramping the gate bias will give us a bias dependent capacitance with negligible current flow. So for an NMOS, we may have the statement

```
SOLVE VGATE=0.0 NAME=GATE VSTEP=0.01
VFINAL=2.0 QSCV
```

to give us the quasi-static C-V curve as it is biased into inversion. It is best to use a fine voltage step with QSCV to give good resolution. The process can be repeated by setting the S-P model in the MODELS statement instead of BQP to obtain the same set of curves for the S-P model. The BQP model is then rerun with different sets of parameters until an acceptable agreement with the curves produced by the S-P model is achieved.

In our simulations, we have used the Quasi-static capacitance voltage curves for comparison with the Schrödinger Poisson equations. Later on, the electron concentration variation with depth was also compared. It was found that calibration against either one of these quantities also gave the best possible calibration when the other quantity was compared. Hence, comparing both was unnecessary.

*D. Post Calibration runs [8]*

After obtaining the parameters for BQP, either the Drift-Diffusion or energy balance (hydrodynamic) equations can be solved as usual. For cases where Lattice Heating is important then LAT.TEMP can be enabled at the same time.

The iteration scheme uses a modified version of BLOCK. Set BLOCK in the METHOD statement (although NEWTON and GUMMEL are ignored, BLOCK is always used if the BQP model is set). If an Energy Balance model is chosen (HCTE.EL or HCTE.HO on the MODELS statement), then an alternative iteration scheme will become available by specifying BQP.ALTEB in the METHOD statement. This method is slower and is only available in ATLAS2D but may have better convergence properties.

By using BQP.NOFERMI, the BQP equation will only use its Boltzmann statistics form. Without this parameter, the statistics used are those specified for the other equations. With fermi statistics, the convergence can be poor for very high carrier densities, and this parameter can circumvent some convergence properties. Re-calibrate the BQP parameters if you set BQP.NOFERMI.

To speed up convergence, we specified the NOCURRENT parameter on the first SOLVE statement after SOLVE INIT. It prevented the need to use the QFACTOR parameter as was necessary for the Density Gradient method. QSCV enables the quasi-static capacitance calculation and output.

## 3D Quantum Simulation Results

In this section, we present the results of the 3D quantum simulation of a silicon nanowire structure.

*A. Device structure*

The simulated device is a three-dimensional structure with gates all around the silicon channel as shown in Fig. 1 and its cutplane is shown in Fig.2. Device simula-tions are performed using Silvaco's 3-D ATLAS device simulation environment. Half of the device is constructed in a 2-D platform and then extended in the z plane to create a 3-D nanowire structure for simulations. The device cross section is 5 nm X 5 nm (rectangular) while its effective channel length is varied between 5 and 100 nm. The channel region is assumed to be p doped ($N_A = 10^{16}$ cm$^{-3}$) and the source and drain extension regions are heavily n-doped ($N_D = 10^{20}$ cm$^{-3}$) with abrupt doping profiles. A metallic gate with the mid-gap work function of 4.7 eV is assumed and the gate oxide is 2 nm thick.

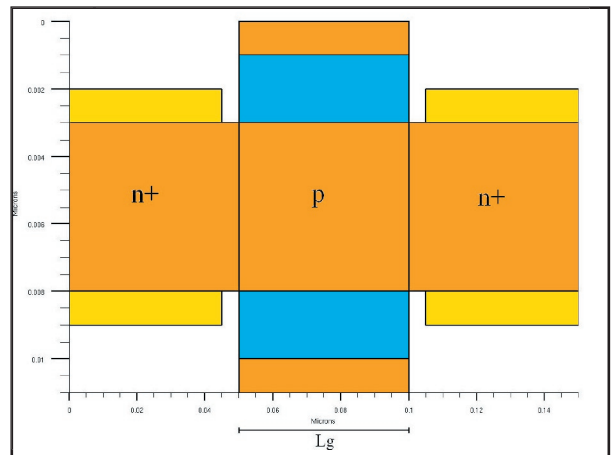

Fig.2 Y-Z cutplane view of the Nanowire.

*B. Quantum Simulation using BQP Model*

As discussed earlier, the Bohm Quantum Potential (BQP) model [9] calculates a position dependent potential energy term using an auxiliary equation derived from the Bohm interpretation of quantum mechanics. This extra potential energy modifies the electron and/or hole distribution. Fig. 3 shows an isosurface of electron Bohm Quantum Potential value of 0.1V and and Fig. 4 shows its cutplane across the center of the channel showing the electron BQP variation. It is observed that there is localization near the perimeters of the device. The effect of the EBQP is to reduce the electron density around the edges of the channel, where EBQP is negative and increase electron density in the center, where EBQP is positive. The 2D electron concentration is plotted in Fig. 5 for 0.5 V bias applied to the gate. It clearly shows electrons are repelled from the interface Si/Oxide in all four directions equally.

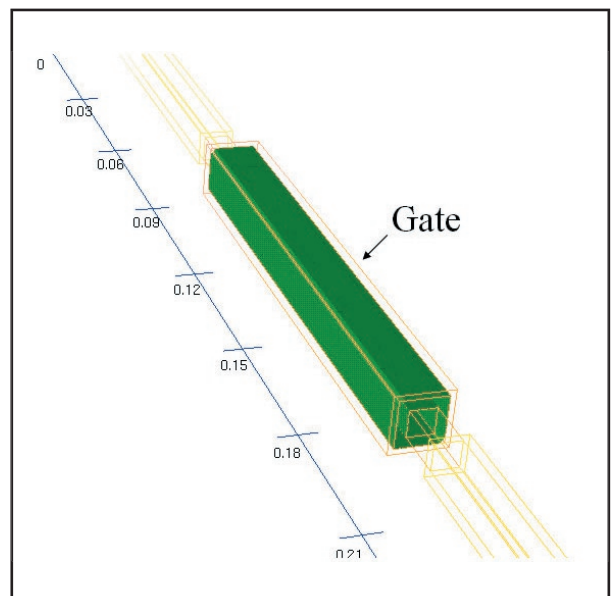

Fig.3 An isosurface of constant electron Bohm Quantum Potential with a value 0.1 V.

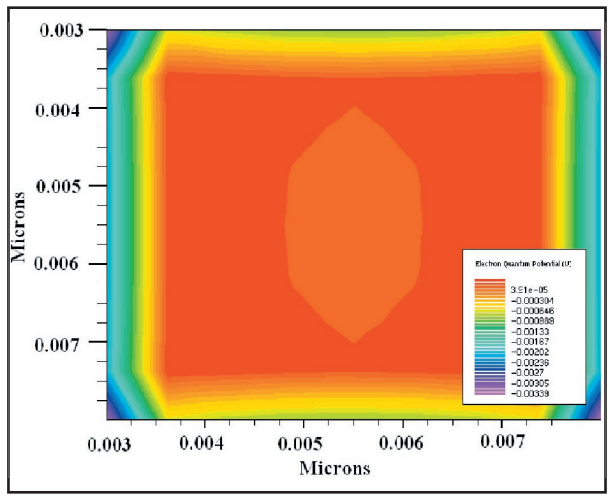

Fig.4 Cutplane across the center of the Silicon Nanowire FET channel showing the 2D electron BQP variation.

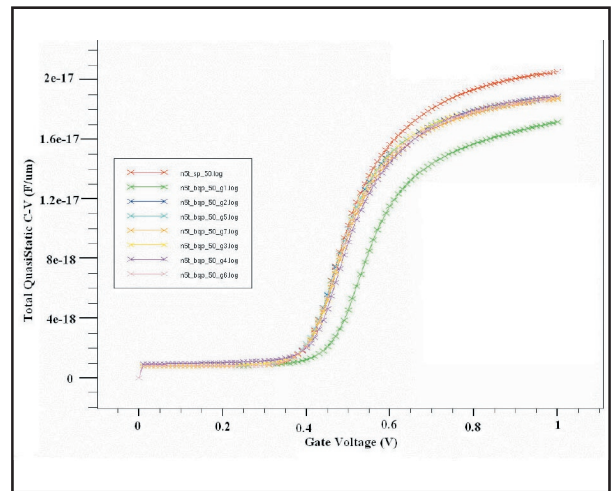

Fig.6 Calibration of BQP parameters for the 50 nm channel length Nanowire FET.

## C. Calibration of BQP model

For calibration of the BQP model with Schrodinger Poisson results, we chose to compare the quasi-static capacitance-voltage curves. In ATLAS, the low frequency C-V curve can be computed in static operation: the charge concentration is integrated in the whole structure and then this quantity is derived as C=-dQ/dV. In Fig. 6 we show the calibration of Bohm Quantum Potential model against the Schrodinger-Poisson model for a NWFET of channel length 50 nm being biased into inversion. The calibration takes place under conditions of negligible current flow. The voltage range was 0-1V volts which includes the depletion to inversion transition region.

First, the Quasi-static CV curves using the Schrodinger-Poisson model were evaluated. Quantum confinement is strong in both the x and y directions, so a complete solution of the 2D Schrodinger equation in two dimensions is required. QSCV is specified on the solve statement and a fine voltage increment is used in order to give accurate calculation of the quasi-static capacitance as a function of voltage. Next, the C-V curves over the same voltage range and with the same voltage increment as for the S-P model are simulated. Multiple different sets of BQP pa-rameters are used for each of these simulations, in order to get a close calibration as shown in Fig. 7 where the closest overlap is obtained for gamma = 0.01 and alpha = 0.3 for the 50 nm channel length nanowire FET. The electron concentration profiles were also compared using the calibrated BQP model values to confirm the calibration as shown in Fig. 8 obtained at $V_{DS}$ = 0.5 V.

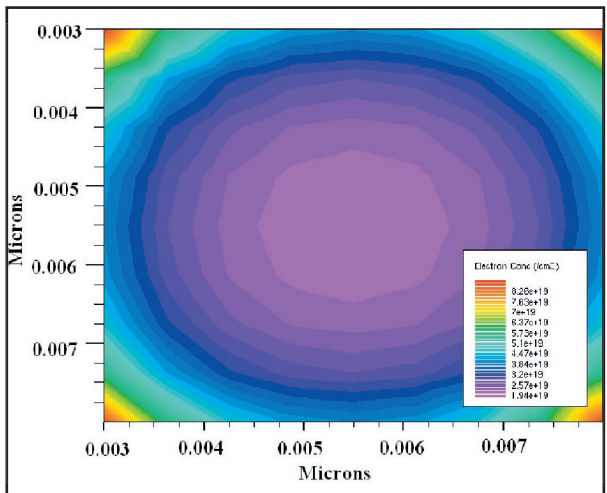

Fig. 5 Cutplane across the center of the Silicon Nanowire FET channel showing the 2D electron concentration variation.

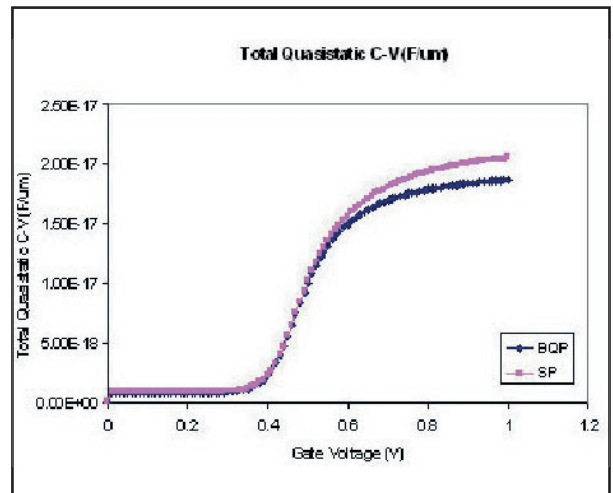

Fig.7 Total quasi-static CV showing the closest match between BQP and SP.

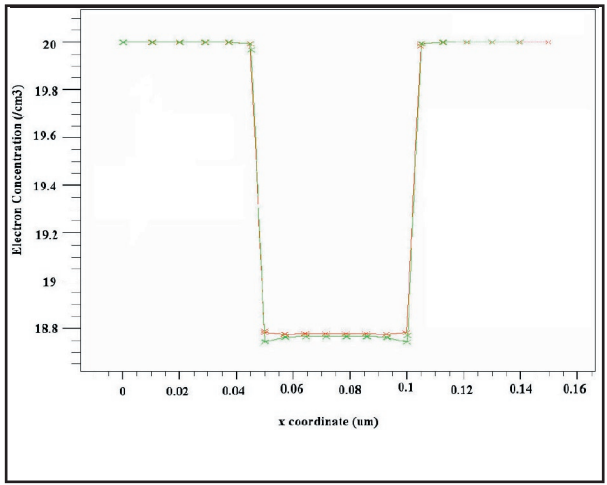

Fig. 8 Plot of electron concentration variation across the channel using the calibrated BQP model.

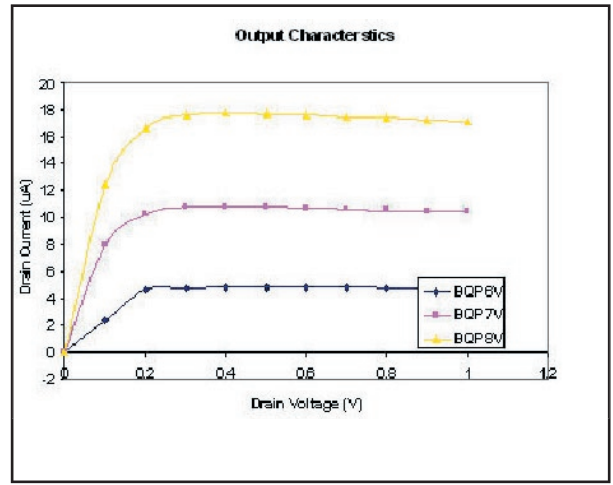

Fig. 10 Output characteristics computed $V_{GS}$ = 0.6, -.7 and 0.8 V for L = 50 nm using BQP model.

As can be observed in Fig. 8, the effective quantum potential method is capable to generate accurate shape of the carrier density profile as the applied voltage is varied. The method is also capable to accurately reproduce the carrier density per unit area obtained with the direct solution of the Schrödinger equation. For the particular device whose calibration is performed (in this case, the 50 nm length NWFET), the calibrated set of BQP parameters are used for all the further calculations.

### D. Device characteristics using calibrated BQP model

The transfer and output characteristics for the device were obtained using the calibrated BQP model. The effect of using the BQP model is shown in Fig. 9 by comparing against the results obtained from semi classical simulation, which does not include quantum confinement effects. The effect of BQP model is to increase the drain current compared to the current obtained when quantum confinement is not included. The output characteristics computed for different gate voltages using the BQP model are shown in Fig. 10.

### E. Channel Length scaling

We studied the scaling behavior of SNWFETs by varying the length of the channel in the device. Fig. 11 shows the variation in transfer characteristics when the channel length is reduced from 100 nm to 10 nm. The cross section of the channel was 5 nm x 5nm in each case. It can be seen that the device characteristics degrade as the channel length is reduced. As the channel length becomes smaller, the drain current does not saturate, as is seen for the case of 10 nm channel length device shown in Fig. 12 for $V_{GS}$ = 0.6 V. This is due to the Drain Induced Barrier Lowering (DIBL).

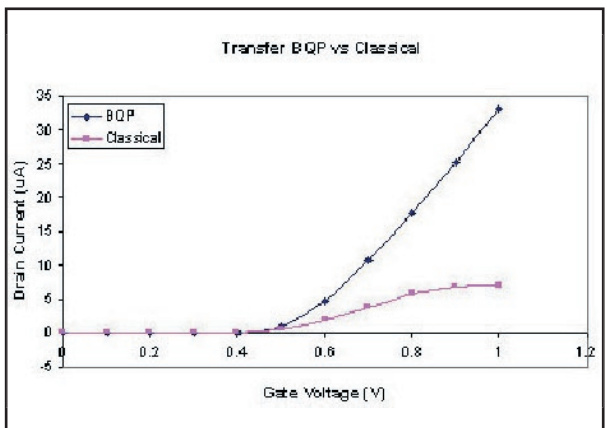

Fig. 9 Transfer characteristics computed for a 50 nm channel lengh nanowire FET using semi-classical and quantum methods (VDS = 0.5 V).

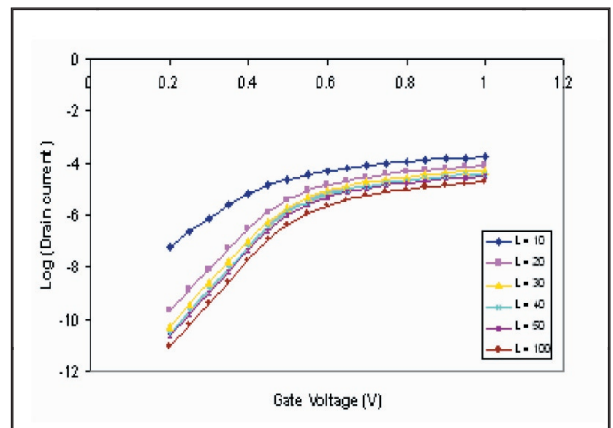

Fig. 11 Transfer characteristics of a Silicon Nanowire FET with a cross section 5 nm × 5 nm and for channel lengths L = 100, 50, 40, 30, 20, 15, and 10 nm.

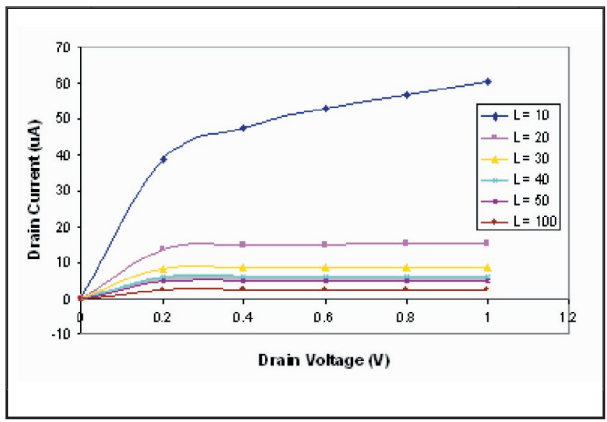

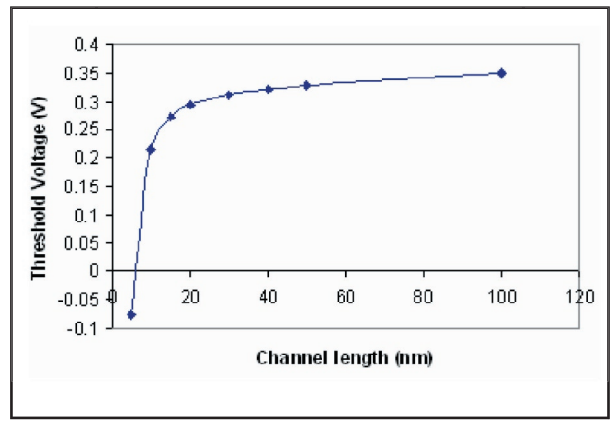

Fig. 12 Output characteristics of a Silicon Nanowire FET with a cross section 5 nm x 5 nm and for channel lengths L = 100, 50, 40, 30, 20, 15, and 10 nm.

Fig. 14 Threshold voltage variation as a function of channel length of a Silicon Nanowire FET with a cross section 5 nm x 5

To understand the DIBL effects, the conduction band energy along the channel from source to drain is shown in Fig. 13 for different drain voltages for a 20 nm channel length nanowire FET. This figure clearly demonstrates the lowering of the barrier potential on the source side as the drain voltage increases. Threshold voltage variation as a function of channel length shown in Fig. 14 also demonstrate the lack of gate control due to short channel effects.

## Conclusions

ATLAS Quantum provides a set of models for simulation of various effects of quantum confinement and quantum transport of carriers in semiconductor devices. We have performed a quantum simulation of Silicon Nanowire Field-Effect Transistors by using the Bohm Quantum Potential model to include quantum confinement effects in the simulations. The BQP model was successfully calibrated against 2D Schrodinger Poisson simulation results under conditions of negligible current flow by matching the Quasi-static Capacitance Voltage curves.

The scaling behavior of the silicon nanowire field effect transistors has been investigated, and it has been found that their device characteristics sensitively depend on the channel length. The device performance has been found to degrade sharply for channel lengths in the sub 20 nm regime.

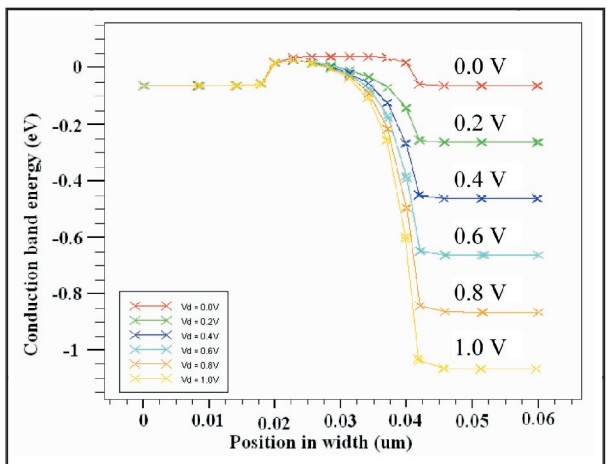

Fig. 13 Conduction band energy at Y=0 as a function of X at different drain voltages $V_{DS}$ = 0.0, 0.2, 0.4, 0.6, 0.8, 1.0 V and $V_{GS}$ = 0.6 V.